\documentclass[sigconf,10pt]{acmart}
\usepackage{booktabs}
\usepackage{algorithm}
\usepackage[noend]{algpseudocode}
\usepackage{algorithmicx}

\usepackage{subcaption}
\usepackage{times}
\usepackage{multicol}
\usepackage{xcolor}

\algdef{SE}[SUBALG]{Indent}{EndIndent}{}{\algorithmicend\ }%
\algtext*{Indent}
\algtext*{EndIndent}

\newcommand{\ignore}[1]{}
\newcommand{\enote}[1]{\begin{quote}{\bf Erez's note:} #1\end{quote}}

\AtBeginDocument{%
  \providecommand\BibTeX{{%
    \normalfont B\kern-0.5em{\scshape i\kern-0.25em b}\kern-0.8em\TeX}}}

\setcopyright{acmcopyright}
\copyrightyear{2018}
\acmYear{2018}
\acmDOI{XXXXXXX.XXXXXXX}

\setcopyright{none}
\begin{document}

\title{The ERA Theorem for Safe Memory Reclamation}

\author{Gali Sheffi}
\email{sheffiga@gmail.com}
\affiliation{%
  \institution{Department of Computer Science, Technion}
  \city{Haifa}
  \country{Israel}
}

\author{Erez Petrank}
\email{erez@cs.technion.ac.il}
\affiliation{%
  \institution{Department of Computer Science, Technion}
  \city{Haifa}
  \country{Israel}
}

\renewcommand{\shortauthors}{Gali Sheffi and Erez Petrank}

\begin{abstract}
Safe memory reclamation (SMR) schemes for concurrent data structures offer trade-offs between three desirable properties: ease of integration, robustness, and applicability. 
In this paper we rigorously define SMR and these three properties, and we present the ERA theorem, asserting that any SMR scheme can only provide at most two of the three properties. 
\end{abstract}



\maketitle

\section{Introduction} \label{sec-intro}
Managing memory for concurrent data structures is known to be non-trivial. The main problem is that a node $n$ that is detached from the data structure and is headed for reclamation, may still be accessed by concurrent threads that gained  access to $n$ prior to its detachment. Once a detached node is reclaimed, the executing threads are in danger of accessing freed memory, potentially causing a system crash, a segmentation fault, or correctness failure~\cite{michael2004hazard,michael2004aba}. This problem can be prevented using a rigid access discipline with locking, but such locking is not typically used because it is often detrimental to performance, and because locking foils the progress guarantee of non-blocking data structures. 

To deal with this problem, SMR schemes were presented. 
The task of an SMR scheme is to either prevent hazardous accesses, or delay the reclamation of nodes that are still accessible by concurrent threads. 
With an SMR scheme, nodes are first manually \emph{retired}, announcing that they are candidates for reclamation (and re-allocation) and the SMR scheme is responsible for determining when a retired node can be safely reclaimed and reused. Retired nodes are typically held in pair-thread retire lists 
~\cite{michael2004hazard,wen2018interval,ramalhete2017brief,DBLP:conf/wdag/SheffiHP21,singh2021nbr,harris2001pragmatic} until they are eligible for reclamation. 

Each SMR comes with some benefits over existing schemes, but also with some disadvantages for the concurrent system. In this paper we focus on three good properties that have been mentioned in prior work: Ease of integration, Robustness, and wide Applicability (ERA, in short). We define these properties formally, and present a theorem, asserting that it is not possible for any SMR scheme to deliver all three qualities. 

Concurrent data-structure implementations usually adhere to some correctness criteria, typically, linearizability~\cite{herlihy1990linearizability},
and provide some progress guarantee, typically lock-freedom or  wait-freedom~\cite{herlihy2011nature,herlihy1991wait}. Interestingly, there is no analogue widely accepted correctness criteria for SMR implementations. 
Various works have phrased ad-hoc correctness conditions~\cite{kang2020marriage,michael2004hazard,balmau2016fast,dice2016fast,meyer2019pointer,meyer2019decoupling}. For example,~\cite{michael2004hazard,ramalhete2017brief,wen2018interval} introduce the terms {\em protection}, or {\em hazardous access}, which are specific to the SMR scheme they present, but are not adequate for other schemes. Some previous work prove ad-hoc properties that relate to correctness~\cite{DBLP:conf/wdag/SheffiHP21,wen2018interval,singh2021nbr}, but there is no correctness condition that is applicable to all SMR techniques in the literature. 
Clearly, a reclamation scheme should preserve the original implementation's correctness (e.g., linearizability) and it is desirable to also keep its progress guarantee. The notion of safety is often used, but with no rigorous acceptable definition available. For example, safety may be interpreted as preventing  access to reclaimed memory, but then, optimistic methods~\cite{DBLP:conf/wdag/SheffiHP21,cohen2015automatic,cohen2015efficient}, that allow access to reclaimed data, may erroneously be considered as unsafe.

In addition to the lack of an adequate definition for safe memory reclamation, there are also some properties of SMR schemes that have never been rigorously defined. Previous works~\cite{singh2021nbr,kang2020marriage,nikolaev2021snapshot} listed various desirable SMR properties, which are often used to evaluate the overall quality of an SMR scheme. The most natural properties that we care about include performance (and scalability), progress guarantee, applicability to a wide set of concurrent data-structure implementations, the ease of integration with a given concurrent data structure, and the memory footprint, a.k.a. robustness. While some of these properties are intuitively clear, most of them were not formally defined. Moreover, robustness was only recently introduced, only in the scope of memory reclamation~\cite{balmau2016fast,dice2016fast}, and is sometimes used to mean different things. Let us briefly discuss the three notions that are at the focus of this paper.  

Defining wide {\em applicability} is not straightforward. It is not always easy to tell if a given SMR scheme is applicable to a given data structure. For example, it is not trivial to see that the Hazard Pointers (HP) method~\cite{michael2004hazard} and some of its extensions~\cite{ramalhete2017brief,wen2018interval,nikolaev2020universal} are not applicable to Harris's linked-list~\cite{harris2001pragmatic}, and it is not clear how to test HP's applicability to a new data structure. While the applicability notion has been considered previously~\cite{singh2021nbr,kang2020marriage,DBLP:conf/wdag/SheffiHP21,cohen2015automatic}, it was never formally defined.   
	

{\em Robustness} has two different meanings in the literature. According to Dice et al.~\cite{dice2016fast}, an SMR scheme is considered as robust if there is a bound on the number of retired objects that cannot be reclaimed. According to Balmau et al.~\cite{balmau2016fast}, the number of reclamation-related computational steps should be bounded.
While the first definition has subsequently been more widely accepted~\cite{singh2021nbr,wen2018interval,DBLP:conf/wdag/SheffiHP21}, both suffer from their informality. First, the bound is not defined. It is unclear whether it should be a constant, or whether it may depend on the size of the data structure, or even the execution length, and in what way. If the data structure requires space, exponential in the number of insert operations, is it robust? If a scheme exhausts the heap even with small data structures, is it robust? Furthermore, what does "cannot be reclaimed" mean? Various SMR schemes are driven by different reclamation triggers, and a reclamation of a certain set of objects may be postponed, regardless of safety. 
Moreover, the terminology in existing definitions is not obvious. What is a reclamation-related step? Some SMR schemes change the original implementation's layout fundamentally, making the difference between the original implementation steps and the newly inserted ones ambiguous. We feel that a formal definition of robustness is missing. The definition should rigorously clarify the concept of memory overhead for SMR schemes.

Finally, we look at how difficult it is to integrate an SMR scheme into a given data structure, i.e., at {\em ease of integration}. 
An SMR scheme typically provides a set of API operations that should be inserted into the given code (e.g., \emph{alloc()}, \emph{retire()}, \emph{beginOperation()}~\cite{harris2001pragmatic,fraser2004practical,wen2018interval}).
Sometimes, the integration also involves changing the original program. E.g., the Automatic Optimistic Access (AOA) scheme~\cite{cohen2015automatic} requires the data structure to be in a normalized form~\cite{timnat2012wait} for the  integration. Free Access (FA)~\cite{Cohen18} and Neutralization-Based Reclamation (NBR)~\cite{singh2021nbr} require that the code is divided into separate read and write phases, and Version-Based Reclamation (VBR)~\cite{DBLP:conf/wdag/SheffiHP21} provides a designated mechanism for adding code checkpoints.
Note that wide applicability and easy integration are independent properties. A scheme can be easily integrated to any given code, but, still, not be applicable to some data-structure implementations. 

In this paper, we formally define safe memory reclamation, along with the three desirable properties: robustness, wide applicability and ease of integration. 
%
In addition, we present and prove the ERA theorem, asserting that the three properties (Ease of integration, Robustness, and wide Applicability) cannot co-exist. I.e., any SMR scheme can have at most two out of the three desirable properties.
%
This paper is organized as follows: Related work is surveyed in Section~\ref{sec-related}. In Section~\ref{sec-background} we specify the shared-memory model. We formally define safe memory reclamation in Section~\ref{sec-backgound-smr}, and the three desirable properties in Section~\ref{sec-properties}. We present the ERA Theorem in Section~\ref{sec-impossibility}, and conclude in Section~\ref{sec-conclusion}.


\section{Related Work} \label{sec-related}

Detailed surveys of SMR schemes appear in the literature~\cite{DBLP:conf/wdag/SheffiHP21,singh2021nbr,brown2015reclaiming,wen2018interval}. We focus on previous efforts to address and define suitable correctness conditions for SMR schemes and desirable reclamation properties.
To the best of our knowledge, while there exist formal methods for verifying safety in manually reclaimed environments~\cite{gotsman2013verifying,meyer2019pointer,meyer2019decoupling}, and various reclamation schemes have been shown to posses highly desirable properties (e.g., robustness, easy integration and wide applicability), these notions have never been formally defined. 

Michael~\cite{michael2004hazard} and Herlihy et al.~\cite{herlihy2005nonblocking} were the first to address the need in bounding the amount of memory occupied by removed elements. 
In both suggested schemes, each thread has a pool of global pointers (called hazard pointers in~\cite{michael2004hazard} or guards in~\cite{herlihy2005nonblocking}), used to postpone the reclamation of certain nodes that might still be in use. Both schemes were designed to provide an upper bound guarantee on the total number of deleted nodes that are not yet eligible for reuse at any given moment.
Braginsky et al.~\cite{braginsky2013drop} suggested a flexible trade-off between space and  runtime overhead suggested by these two reclamation schemes.
The term \emph{Robustness} was first introduced by Dice et al.~\cite{dice2016fast} and Balmau et al.~\cite{balmau2016fast}. While the first definition put a limit on the number of deleted  objects that could not be reclaimed, the latter bounded the number of steps during any action related to memory reclamation.
Subsequently, the first definition (bounding the space overhead) was widely adopted~\cite{ramalhete2017brief,wen2018interval,kang2020marriage,singh2021nbr,nikolaev2020universal,nikolaev2021snapshot,DBLP:conf/wdag/SheffiHP21}. However, the bound on the space overhead was never specified, and so reclamation schemes with liberal bound (that can potentially exhaust the available memory)~\cite{ramalhete2017brief,wen2018interval,nikolaev2020universal} are considered robust.
Reference counting-based schemes~\cite{gidenstam2008efficient,detlefs2002lock,herlihy2005nonblocking} are usually not robust, mainly due to the existence of cyclic structures of retired objects (these cycles can be broken~\cite{sundell2005wait,correia2021orcgc}, at the cost of higher performance overheads).

Although some reclamation schemes are designed for a limited set of  data-structures~\cite{braginsky2013drop}, most methods are designed for a general, wide set of data-structure implementations.
A common issue which rises in this context is that many allegedly general schemes~\cite{michael2004hazard,ramalhete2017brief,wen2018interval,nikolaev2020universal,solomon2021efficiently} require restricted access to deleted nodes. I.e., traversing a deleted node is forbidden. This requirement is problematic, as many state-of-the-art lock-free data-structures allow such traversals~\cite{heller2005lazy,harris2001pragmatic,natarajan2014fast,brown2014general,ellen2010non} to obtain fast searches. 
Indeed, this issue was discussed in various works~\cite{cohen2015automatic,fraser2004practical,kang2020marriage,brown2015reclaiming}. An extensive study by Singh et al.~\cite{singh2021nbr} lists many popular lock-free data-structures for which such schemes are not applicable. 
Michael~\cite{michael2002high} suggested a modification of Harris's lock-free linked-list~\cite{harris2001pragmatic}, which makes it suitable for these reclamation schemes. However, this modification reduces the performance of the linked-list (for more details, see~\cite{cohen2015automatic}). Furthermore, it is not a universal construction and it cannot be used to modify other data-structures.
Gidenstam et al.~\cite{gidenstam2008efficient} suggested a more general solution, but their construction adds even higher performance overheads, and involves a complicated manual integration into the given code.
Wide applicability was also referred to as \emph{generality} in~\cite{nikolaev2020universal}.

Applicability does not guarantee an easy integration. First, the programmer must issue \emph{retire()} statements at a location where the node is already detached from the data structure. 
Also, reclamation schemes often require additional integration. The simplest integration is provided by the seminal EBR scheme~\cite{harris2001pragmatic,fraser2004practical,brown2015reclaiming}, which solely requires inserting calls to external methods in the beginning and end of each data-structure operation.
Other schemes~\cite{michael2004hazard,wen2018interval,ramalhete2017brief,kang2020marriage} provide a slightly more complicated interface, which is still relatively easy for integration. Such interfaces may include an explicit node protection method (preventing the reclamation of retired nodes while they are potentially still in use), and read and write barriers (i.e., code to be executed with any read or write of a data structure field). Such methods can be automatically integrated into an existing implementation, and do not require a significant familiarity with the original code. 
However, some reclamation schemes (e.g.,~\cite{cohen2015automatic,cohen2015efficient,brown2015reclaiming,singh2021nbr,DBLP:conf/wdag/SheffiHP21}) require a more complicated integration procedure. Such schemes may cause the original algorithm to fail in accessing data structure fields and retry. This implies control flow changes to determine where to branch when such a failure occurs. Such integration is non-trivial, and it requires a deep understanding of the underlying data structure. Automatic integration is no longer possible.   
However, as the ERA Theorem asserts (see Section~\ref{sec-impossibility}), such harder integration efforts are required to obtain general applicability and a low space overhead.

In addition to performance, robustness, wide applicability, and easy integration, other reclamation properties have been considered in previous work.
Singh et al.~\cite{singh2021nbr}, considered \emph{consistency}, which requires the overall performance to not be affected by workload changes or under a system over-subscription.
Nikolaev and Ravindran~\cite{nikolaev2021snapshot} claimed that a reclamation scheme should be \emph{transparent}. Meaning, threads can be created and deleted dynamically throughout the execution, and without affecting the scheme's safety.
Cohen and Petrank introduced an \emph{optimistic} reclamation method~\cite{cohen2015automatic,cohen2015efficient}, which allows threads to read reclaimed memory, while taking care to preserve correctness. Such schemes provide strong robustness and high throughput. Sheffi et al.~\cite{DBLP:conf/wdag/SheffiHP21} extended optimistic access to writes, providing a fully optimistic solution.

In addition to ensuring desirable properties, avoiding common unwanted side affects is also important.
Many reclamation schemes introduce significant memory overheads, as they add extra fields to the data-structure nodes' layout. E.g., epoch-related data~\cite{DBLP:conf/wdag/SheffiHP21,wen2018interval,ramalhete2017brief}, or reference counters~\cite{gidenstam2008efficient,detlefs2002lock,herlihy2005nonblocking}.
Optimistic methods rely on type-preservation~\cite{cohen2015automatic,cohen2015efficient,DBLP:conf/wdag/SheffiHP21}, which means that nodes should be re-allocated to the same type. This is adequate particularly when there is a small number of major data structures in the code. 
Some reclamation algorithms require special compiler or hardware support (and therefore, are not considered as \emph{self-contained}~\cite{kang2020marriage}). E.g., DEBRA+~\cite{brown2015reclaiming} and NBR~\cite{singh2021nbr} rely on lock-free OS signals to preserve lock-freedom, VBR~\cite{DBLP:conf/wdag/SheffiHP21,sheffi2021vbr} and Hyaline~\cite{nikolaev2020universal} rely on a hardware-provided wide CAS instruction, StackTrack~\cite{alistarh2014stacktrack} and ThreadScan~\cite{alistarh2018threadscan} rely on transactional memory, Dice et al.~\cite{dice2016fast} and PEBR~\cite{kang2020marriage}  rely on the existence of process-wide memory fences, and QSense~\cite{balmau2016fast} requires control over the OS scheduler.

\section{Preliminaries} \label{sec-background}


We follow previous work and use the basic asynchronous shared memory model from~\cite{herlihy1991wait}, and related definitions from~\cite{herlihy1990linearizability}. 
We consider a fixed set of $N$ executing threads, communicating by applying operations on \emph{shared objects}.
An object is an instance of an \emph{abstract data type}, which specifies a set of possible values, and a set of operations that provide the only means to access it. 
For example, we define the \emph{set} data type as a set of integer keys (denoted as \emph{set keys}), initially empty. Its associated operations are \emph{insert(key)}, \emph{delete(key)} and \emph{contains(key)}, where \emph{key} is an integer. The \emph{insert(key)} operation inserts \emph{key} into the set and returns \emph{true} if the set does not already contain \emph{key}, and returns \emph{false} otherwise. The \emph{delete(key)} operation removes \emph{key} from the set and returns \emph{true} if the set indeed contains \emph{key}, and returns \emph{false} otherwise. The \emph{contains(key)} operation returns \emph{true} if the set contains \emph{key}, and returns \emph{false} otherwise. 

\paragraph{Implementations and Data-Structures}
An \emph{implementation} of an object (or several objects) provides a data-representation by applying primitive memory access operations (e.g., reads, writes, atomic read-modify-write instructions~\cite{herlihy1991wait}) on a set of base objects (i.e., shared memory locations).
Specifically, \emph{set objects} are represented by \emph{shared data-structures}. Each set key is represented by a \emph{node}, and each data-structure has a fixed set of entry points (e.g., a linked-list head~\cite{harris2001pragmatic} or a tree root~\cite{natarajan2014fast}), which are node pointers. 
Nodes may contain node pointer fields. We say that a node $m$ is a successor of a node $n$ (or that $n$ is a predecessor of $m$) if at least one of $n$'s node pointer fields points to $m$.
Accordingly, we say that a node $m$ is reachable from a node $n$ if there exist nodes $n_0,n_1,\ldots,n_k$ such that (1) $n=n_0$, (2) $m=n_k$, and (3) for every $0 \leq i < k$, $n_i$ is a predecessor of $n_{i+1}$.
We say that a certain node is \emph{reachable} if it is reachable from an entry point.

\paragraph{Executions}
A \emph{step} is either a shared-memory access, a local variable access, an operation invocation, or the return from an operation. In all cases, the step includes the executing thread id, the accessed object (when exists), and the respective access input and output values. Steps are considered to be atomic.
A \emph{configuration} specifies the
value of each shared memory address and the state of each thread (including the content of its local variables and program counter). The initial configuration $C_0$ is the configuration in which all memory addresses have their initial values and all threads are in their initial states. In particular, all data-structures are initialized, and represent empty sets.
An \emph{execution} $E=C_0 \cdot s_1 \cdot C_1 \cdot  \ldots$ is an alternating sequence of configurations and steps, starting from the initial configuration. Specifically, an \emph{execution of an implementation} is an execution where, starting from the initial configuration, each step is issued according to the given implementation, each memory read matches the preceding configuration, and each memory write is reflected in the following configuration.
Given a sub-sequence $E' = C_k \cdot s_{k+1} \cdot  \ldots \cdot s_m \cdot C_m$, we say that $E'$ is a \emph{solo-run} if the steps $s_{k+1}, s_{k+2}, \ldots, s_m$ are executed by the same thread.


\paragraph{Histories}
An execution is modeled by its \emph{history}, which is its sub-sequence of operation invocation and response steps.
Given an implementation, its set of \emph{derived histories} is the set of all histories that model executions of that implementation.
Given a history $H$ and a thread $T$, we denote with $H|T$ the sub-history of $H$, consisting of exactly all the steps executed by $T$ in $H$. Similarly, given a history $H$ and a shared object $O$ (may be a memory word or a data-structure as described above), we denote with $H|O$ the sub-history of $H$, consisting of exactly all the steps executed on $O$ in $H$.
Accordingly, given a history $H$, a thread $T$ and a shared object $O$, we denote with $H|\langle T,O \rangle$ the sub-history of $H$, consisting of exactly all the steps executed by $T$ on $O$ in $H$.
Two histories $H,H'$ are \emph{equivalent} if for every thread $T$, it holds that $H|T=H'|T$.

Given a history $H$ and an object $O$, we say that $H|O$ is sequential if it begins with an invocation step, and each invocation step (except for possibly the last one) is immediately followed by its matching response.
We say that a history $H$ is a \emph{sequential history} if for every object $O$, $H|O$ is sequential.
An object is associated with a
\emph{sequential specification}, which is a prefix-closed set of all of its possible sequential histories.

Given a history $H$ that contains an operation invocation, we say that this operation is \emph{complete} in $H$ if $H$ also contains its matching response. Otherwise, we say that this operation is \emph{pending} in $H$.
A history is \emph{complete} if all of its contained operations are complete.

\paragraph{Well-Formed Histories}
The standard definition of well-formed histories~\cite{herlihy1990linearizability} assumes that, given a history $H$ and a thread $T$, $H|T$ is a sequence of operation invocations and their immediate matching responses.
However, describing the integration of a safe memory reclamation scheme into a given data-structure implementation requires nesting operations.
I.e., the reclamation scheme's operations (e.g., \emph{retire()}, \emph{alloc()}) are called in the scope of the data-structure operations.
Therefore, we cannot use the standard definition of well-formed histories from~\cite{herlihy1990linearizability}.
Instead, we follow the extended definition from~\cite{attiya2018nesting}.
Given a history $H$ and an object $O$, we say that $H|O$ is \textit{well-formed} if for every thread $T$, $H|\langle T,O \rangle$ starts with an invocation step, and is an alternating sequence of invocation steps and their immediate matching response.
We say that a history $H$ is \textit{well-formed} if (1) for every object $O$, $H|O$ is well-formed, and (2) for every thread $T$, two of its invocation steps $s_{inv_1}, s_{inv_2}$ and their respective matching response steps $s_{res_1}, s_{res_2}$, if $s_{inv_1}$ precedes $s_{inv_2}$ and $s_{inv_2}$ precedes $s_{res_1}$ in $H$, then $s_{res_2}$ precedes $s_{res_1}$ in $H$.
A \emph{well-formed implementation} is an implementation for which all derived histories are well-formed.

\paragraph{Linearizability}
A complete history $H$ is linearizable if it is well-formed, and for every object $O$, its sequential specification contains a sequential history $S$ such that (1) $H|O$ and $S$ are equivalent, and (2) if a response step precedes an invocation step in $H|O$, then it also precedes it in $S$.
A history $H$ is linearizable if it can be completed (by adding matching response steps to a subset of pending operations in $H$, and removing the rest of $H$'s pending operations) to a linearizable complete history.
A \emph{linearizable implementation} is an implementation for which all derived histories are linearizable.

\paragraph{Lock-Freedom}
We follow Herlihy and Shavit's definition of lock-freedom~\cite{herlihy2011nature}.
Given an execution $E=C_0 \cdot s_1 \cdot \ldots$ and an executing thread $T$, we say that $T$ is \emph{effective} in a configuration $C_i$ if $T$ performs the step $s_j$ for some $j > i$ (informally, $T$ is not starved by the scheduler).
Now, let $s_k$ be an operation invocation by a thread $T$ during an execution $E$. We say that $s_k$ is \emph{effective} if either $s_k$ has a matching response step in $E$, or $T$ is effective in $C_m$ for every $m \geq k$.

A history $H$ provides \emph{minimal progress} if in every suffix of $H$, some pending effective invocation has a matching response. $H$ provides \emph{maximal progress} if in every suffix of $H$, every pending effective invocation has a matching response.
An implementation is \emph{lock-free} if every respective history provides minimal progress, and some respective history provides maximal progress.

%




\section{Defining Safe Memory Reclamation} \label{sec-backgound-smr}

In this section we present a formal definition of safe memory reclamation.
For ease of presentation, we focus on data-structures that implement set objects. This allows defining a life cycle of a node in the data structure. Extensions to other object types are not difficult.

\subsection{Nodes' Life-Cycles} \label{sec-life-cycle}
Following Meyer and Wolff~\cite{meyer2019pointer,meyer2019decoupling} we assume that (for a set implementation) each node goes through stages in a life-cycle, and can be in one of four possible states: unallocated, local, shared, or retired.
Initially, a node is \emph{unallocated}. I.e., its memory is not available for use by the executing threads. After being allocated by a certain thread, the node becomes \emph{local}. While being local, no thread but the allocating thread has access to this node. In particular, the node cannot be reachable (from an entry point of the data structure), and it cannot represent a set item at this stage. Next, the node may or may not become \emph{shared} (e.g., by making it reachable). While being shared, the node may become alternately reachable and unreachable, and may also represent a set item. When a node is either local or shared, we also say that it is \emph{active}.
At some point, an executing thread may retire the node, announcing that this node is about to become garbage. Once a thread retires the node, it becomes \emph{retired} (and cannot be retired again).
Note that some nodes never become shared, and therefore become retired after being local. 
In the lifetime of a node, we assume that a node always becomes unreachable before it becomes retired (generally, nodes can only be reachable while they are shared\footnote{This assumption is necessary for most safe memory reclamation schemes. However, there exist scenarios in which it is not needed~\cite{wei2021constant}.}).
Finally, a retired node may be reclaimed, meaning that its memory may now be used for re-allocation and its state becomes \emph{unallocated} again. We consider nodes as logical entities. I.e., after a node returns to being unallocated, a new allocation from the same address is considered as an allocation of a different node (even if both nodes are allocated on the same address).

\subsection{Safe and Unsafe Memory Accesses}\label{sec:safe-access}

In this section we define a safe memory access.
We think of the memory as segregated into two separate spaces -- the \emph{system space} and the \emph{program space}. The program space is the area in which the executing threads keep their local and shared nodes and variables. In particular, new allocations are always within the program space, and all nodes reside in the program space, until they are reclaimed. At the time of reclamation, the memory reclamation scheme decides whether to keep the reclaimed nodes for potential subsequent re-allocation in the program space, or to return the node space to the system, in which case, the node moves to system space. If the program attempts to access a node in system space, the result is undefined, and may include a segmentation fault. 

From now on, we refer to a given set implementation (with no memory reclamation) as the \emph{plain} implementation. Note that although plain implementations do not include memory reclamation, they do follow the life cycle defined in Section~\ref{sec-life-cycle}. Specifically, they include adequate \emph{retire()} instructions.
Upon integrating a given memory reclamation scheme into a plain implementation, we refer to the derived implementation as the \emph{integrated} implementation. We further use the \emph{plain} and \emph{integrated} terms to describe the respective executions and histories.

{\em Dereferencing} a pointer $p$ means reading or updating a value in a node whose address is stored in the pointer $p$.\footnote{Note that the pointer content is not necessarily equal to the stored address (e.g., marked pointers~\cite{harris2001pragmatic} contain addressees, possibly along with a marked bit).}
After memory is reclaimed, dereferencing a pointer to it might cause a segmentation fault~\cite{michael2004aba}. 
Given an execution $E=C_0 \cdot s_1 \cdot \ldots$ , a pointer variable $p$, and any configuration $C_m$ in the execution ($m \geq 1$), let $s_i$ be the last update of $p$ in the sub-execution $C_0 \cdot s_1\cdot \ldots C_m$. Namely, for $i\leq m$, $p$ is updated in $s_i$, and for every $i<j\leq m$, $s_j$ does not update $p$.
The update of $p$ in $s_i$ may be an allocation of a new node to $p$ or a pointer assignment to $p$ from another pointer $q$.\footnote{In the following clean formalization, we assume no address arithmetic, and that a pointer points to the head of the object. One can easily extend the definitions below to programs that use pointer arithmetic, as long as there is a clear mapping from pointers to nodes.} 

Let $n$ be the node that is referenced by $p$ in $C_m$. We separate into two cases according to whether $s_i$ is an allocation or an assignment. If the last update of $p$ is an allocation in $s_i$, and the node is not in an \emph{unallocated} state in any of the intermediate configurations $C_j$ (for $i \leq j \leq m$),  then we say that $p$ is \emph{valid} in $C_m$. Otherwise, we say that $p$ is \emph{invalid} in $C_m$. For example, by definition, $p$ is always valid in $C_i$.
The second case is that $s_i$ is a pointer assignment, and let $q$ be the pointer whose content is assigned into $p$ in $s_i$. Similarly, we say that $p$ is \emph{valid} in $C_m$ if $q$ is valid in $C_i$, and for all intermediate configurations $C_j$ for $i \leq j \leq m$, $n$ is not in an \emph{unallocated} state in $C_j$. If $p$ is not valid in $C_m$, we say that $p$ is \emph{invalid} in $C_m$.
We can now formally define a safe memory accesses. 

\begin{definition} \label{definition-safe-access}
A memory access is unsafe if it dereferences an invalid pointer. It is safe otherwise. 
\end{definition}

\subsection{Defining SMR in the Presence of Unsafe Memory Accesses}

When designing a memory reclamation scheme, one must either provide an adequate solution for coping with unsafe memory accesses, or make sure that all memory accesses are safe.
Some SMR schemes allow unsafe memory accesses, while taking care to preserve correctness. 
E.g., AOA~\cite{cohen2015automatic} and VBR~\cite{DBLP:conf/wdag/SheffiHP21} allow reading via invalid pointers, as it is ensured that stale values are always ignored. VBR further allows trying to update the shared memory via invalid pointers, as it is guaranteed that the update fails.
Definition~\ref{definition-smr} below encapsulates the conditions for a memory reclamation scheme to be considered as a safe one. Loosely speaking, an SMR should either only use safe memory accesses, or be very careful in how it executes unsafe memory accesses. In particular, it should not access system space (that might end up in a segmentation fault), it should not modify the data on the node (which might have been reclaimed), and it should not use a value read during an unsafe memory access. 

\begin{definition} \label{definition-smr}
A memory reclamation scheme is a safe memory reclamation (SMR) scheme with respect to a given plain implementation, if for each respective integrated execution $E$, all memory accesses in all steps are safe, or if all unsafe memory accesses satisfy the following conditions. Let $s_i$ be a step with an unsafe memory access to a memory node $n$ via a pointer $p$, then the following three conditions must hold:
\begin{enumerate}
    \item In configuration $C_{i-1}$, $n$'s occupied memory belongs to the program space.  
    \item $s_i$ does not update $n$'s content. Namely, $n$'s content in $C_{i-1}$ equals $n$'s content  in $C_i$.
    \item If data from the dereferenced node $n$ is read into a variable or field $v$ (local or shared), then the value in $v$ is never used. \label{C-used}
\end{enumerate}
\end{definition}

The term "used" in Condition~\ref{C-used} refers to the standard program analysis terminology. In particular, if the modified variable (or field) is read in a step $s_j$ (for some $j >i$), then there exists $i < k < j$ such that $s_k$ overwrites the content of this variable (or field).

Note that we only define safety with respect to a given plain implementation, as most schemes are not necessarily safe when integrated with all existing plain implementations.
E.g., the HP~\cite{michael2004hazard} scheme is safe with respect to Michael's linked-list~\cite{michael2002high}, but is not safe with respect to Harris's linked-list~\cite{harris2001pragmatic}. This implies that HP is not applicable to Harris's linked-list plain implementation.
We further discuss applicability (and HP's applicability in particular) in Section~\ref{sec-properties-applicability} and in
Appendix~\ref{section-incompatibility}.

\section{Desirable SMR Properties} \label{sec-properties}

In this section we present formal definitions for three of the most desirable reclamation scheme properties. Robustness is defined in Section~\ref{sec-properties-robustness}, easy integration is defined in Section~\ref{sec-properties-easy-integration}, and wide applicability is defined in Section and~\ref{sec-properties-applicability}.
One property may affect another in a design. For example, robustness may affect progress, and preserving progress guarantees may affect applicability. 
But in the definitions, we separate the notions and define each of them independently of the others. 

\subsection{Robustness (Memory Footprint)} \label{sec-properties-robustness}
Similarly to~\cite{singh2021nbr,DBLP:conf/wdag/SheffiHP21,wen2018interval}, we adopt the definition of robustness from Dice et al.~\cite{dice2016fast}, which relates to the space overhead or memory footprint of a reclamation scheme. 
According to this definition, a reclamation scheme is considered robust if a failed or delayed thread cannot totally prevent memory reclamation.
%
%
This is formalized by a bound on the amount of retired nodes that exist at any point in the execution. As in Section~\ref{sec-life-cycle}, retired nodes are nodes that have already been retired, are not in the state of \emph{unallocated}. This includes nodes that cannot be reclaimed (due to some reclamation condition that the nodes do not satisfy), together with the retired nodes that have simply not yet been reclaimed, typically because some periodic process that reclaims objects has not yet processed them%
~\cite{singh2021nbr,fraser2004practical,brown2015reclaiming}.\ignore{\footnote{Note that one can design a plain implementation in which unlinked nodes are never retired, making all reclamation schemes seem vacuously robust. 
However, robustness is required with respect to all possible plain implementations, including ones that do retire objects.
}
}

To the best of our knowledge, previous work does not specify any general definition for the bound on the number of such objects. It is not clear whether this bound should be a pre-defined constant, may depend on the specific execution, or may depend on the data-structure size. In definitions~\ref{definition-robustness}-\ref{definition-weak-robustness} below we classify the different levels of robustness, according to the bound that an SMR scheme satisfies. 

\ignore{
The strongest requirement that we make is that the number of retired nodes is bounded by a linear function in the number of threads $N$. 

\enote{Definition~\ref{definition-full-robustness} is missing a discussion. Let's discuss what to discuss... Maybe the following.}
\enote{In the following definitions we consider a bound on the number of retired objects that depends on the size of the data structure. The size of the data structure is dynamic and is represented by  }

\begin{definition} \label{definition-full-robustness} \textbf{(Strong Robustness)}
We say that a reclamation scheme is \emph{strongly robust} if there exists a constant $c$ such that for every integrated execution $E=C_0 \cdot s_1 \cdot \ldots$ , the number of retired nodes in configuration $C_i$ is bounded by $c \cdot N$, for all $i\geq 1$.
\end{definition}

Note that $c$ does not depend on the specific execution or implementation. It is universal for all executions of all implementations. 
VBR~\cite{DBLP:conf/wdag/SheffiHP21} is strongly robust, with $c$ being a pre-determined local retired list size.
}

In the following definitions we consider a bound on the number of retired objects that depends on the size of the data structure. The size of the data structure is dynamic and is bounded by the number of active nodes, i.e., the nodes that have been allocated and not yet been retired (see Section~\ref{sec-life-cycle}).

We first define the class of robust reclamation schemes. Robustness bounds the number of retired nodes at any time by a function that is asymptotically smaller than the maximum size of the data structure so far in the execution, multiplied by the number of threads. Formally, given an execution $E$ , we denote the number of active nodes in $C_i$ by $active_E(i)$. In addition, we set the function $max\_active_E(i)$ to be $max \{ active_E(0), \ldots,active_E(i) \}$.


\begin{definition} \textbf{(Robustness)} \label{definition-robustness}
We say that a reclamation scheme is \emph{robust} if for every integrated execution $E$ , there exists a function $f_E : \mathbb{N} \rightarrow \mathbb{N}$, such that (1) $f_E = o(max\_active_E)$, and (2) for every configuration $C_i$, the number of retired nodes in $C_i$ is bounded by $f_E(i) \cdot N$.
\end{definition}

VBR~\cite{DBLP:conf/wdag/SheffiHP21} is robust, with $f_E(i)$ being a constant function, bounded by the local retire list size (which does not depend on the execution). 
This scheme presents the strongest robustness available by an SMR today. Its bound does not depend on the size of the data structure. 

HP~\cite{michael2004hazard}, AOA~\cite{cohen2015automatic}, and NBR \cite{singh2021nbr} all use hazard pointers~\cite{michael2004hazard} for write protection\footnote{The HP scheme uses hazard pointers for read protection as well.}. For all three schemes, $f_E$ depends on the pre-defined local retire list size (similarly to VBR) plus the number of hazard pointers. 
The number of hazard pointers is typically a small constant (e.g., 3 for linked-lists~\cite{michael2002high,harris2001pragmatic}), but may also depend on the number of active nodes (e.g., for skip lists with a dynamic number of levels~\cite{aksenov2020splay}). While the number of hazard pointers is not guaranteed to be asymptotically smaller than the number of live objects, in 
all known data structures it is. It is an open question for the study of hazard pointers to bound their number. If the number of hazard pointers is always asymptotically smaller than the number of active nodes, then all three schemes are robust.

%


Some reclamation schemes do not provide robustness, but a bound still exists. Accordingly, we define a relaxed term of robustness, denoted \emph{weak robustness}, in Definition~\ref{definition-weak-robustness} below. Note that robust schemes are also considered as weakly robust schemes, but not vice-versa.

\begin{definition} \label{definition-weak-robustness} \textbf{(Weak Robustness)}
We say that a reclamation scheme is \emph{weakly robust} if for every integrated execution $E$, there exists a function $f_E : \mathbb{N} \rightarrow \mathbb{N}$, such that (1) $f_E$ is polynomial in $max\_active_E$, and (2) for every configuration $C_i$, the number of retired nodes in $C_i$ is bounded by $f_E(i) \cdot N$.
\end{definition}

A weakly robust scheme might incur a larger space overhead. Usually, this happens only at worst-case scenarios, but a large space overhead is theoretically possible. 
In some hybrids of the epoch-based~\cite{harris2001pragmatic,fraser2004practical} and pointer-based~\cite{michael2004hazard} reclamation approaches, the execution is divided into epochs, and the number of retired nodes is bounded by the number of active nodes during a certain set of epochs. 
Given an integrated execution $E$ and an epoch $e$, that starts in a configuration $C_i$, the total number of active nodes during $e$ is $active_E(i)$ plus the number of allocations during $e$.
In IBR~\cite{wen2018interval}, each thread might prevent the reclamation of nodes that were active during a small set of \emph{reserved} epochs\footnote{There is a trade-off between IBR's easy integration and the bound on the number of reserved epochs. For more details, see Section~\ref{sec-properties-easy-integration}.}. As IBR allows only a constant number of allocations per epoch, the number of retired nodes in a configuration $C_i$ is linear in $max\_active_E(i) \cdot N$ (which is not asymptotically smaller than $max\_active_E$). Therefore, IBR is weakly robust.
%
%
\ignore{
\begin{definition}
We say that a reclamation scheme is not robust if it is not weakly robust.
\end{definition}
}
EBR~\cite{harris2001pragmatic,fraser2004practical,brown2015reclaiming} is not even weakly robust. Once a thread is halted, all subsequently allocated nodes can never be reclaimed. 


\subsection{Easy Integration} \label{sec-properties-easy-integration}

We assume that the plain implementation already contains proper \emph{retire()} invocations. I.e., we do not consider \emph{retire()} calls installations as part of the reclamation scheme integration. 
The obvious easiest integrate-able SMR is the EBR scheme. EBR provides a \emph{retire()} implementation, along with two API operations, \emph{beginOp()} and \emph{endOp()}, to be respectively inserted in the beginning and end of every data-structure operation. I.e., any plain implementation can be easily integrated with EBR, and the integration process does not require an understanding of the plain implementation. A definition of easily integrated scheme should obviously include EBR, but other schemes are also not that hard to integrate. 

Other examples for easily integrated reclamation methods are the HP~\cite{michael2004hazard} and IBR~\cite{wen2018interval} schemes. Both schemes provide designated \emph{alloc()} and \emph{retire()} implementations, along with code to replace reading and writing from shared memory. IBR also provides \emph{beginOp()} and \emph{endOp()} implementations, to be inserted at the beginning and end of each operation, respectively.
Although the integration of both schemes is slightly more complicated than EBR's, they are still considered as easily integrated, as their integration does not require any familiarity with the original code. 
Note that the IBR authors provide a weakly robust IBR variant (discussed in Section~\ref{sec-properties-robustness}), which cannot by easily integrated according to Definition~\ref{definition-easy-integration} below, as this scheme requires inserting roll-back instructions (returning to a previous point in the code) for maintaining a small robustness bound.
\ignore{
Other examples for relatively easily integrated schemes are HP~\cite{michael2004hazard} and HE~\cite{ramalhete2017brief}. Allegedly, these two schemes require some extra effort, as it seems that their integration requires to determine whether certain pointers are no longer in use (this topic is not fully discussed in~\cite{michael2004hazard} and~\cite{ramalhete2017brief}). Their ease of integration properties remain as interesting open questions.
}

Some schemes cannot be easily integrated with many plain implementations. An important example of integration obstacle is the requirement of an SMR to insert roll-back instructions for handling unsafe memory accesses. Namely, when some validation test fails, one needs to return program control to a point in the code from which it is safe to re-execute. Roll-backs are something that we rule out for easy integration. Another undesirable property of an SMR is that it modifies fields of the data structure. An SMR may add fields to a data structure node to be used for its own activity, but expect the SMR to not modify fields that the plain implementation uses. This {\em modularity, encapsulation of information}, and {\em separation of concerns} between the data structure and the SMR activity are standard principles in software engineering. We stress that marking a pointer to signify a deleted object as in Harris' linked-list is not a problem, because this is a modification by the data structure to support its own 
deletion activity, that the SMR is not involved in. In contrast, if the SMR modifies a data structure field, then the programmer that integrates the SMR into the data structure must have an intimate acquaintance with the fields of the data structure to make sure that the SMR does not foil the data structure operations correctness of performance. Such an intimate knowledge, if required during integration, makes the integration difficult. 

Some schemes deal with difficult roll-backs by adhering to specific code shapes, which allow easier placement of roll-back mechanisms. AOA~\cite{cohen2015automatic} requires that the plain implementation is first transformed into a normalized form~\cite{timnat2014practical}. NBR~\cite{singh2021nbr} requires that the code is first divided into separate read and write phases (to be further discussed in Section~\ref{sec-properties-applicability}), which enable easier rollback call installations. 
VBR~\cite{DBLP:conf/wdag/SheffiHP21} relies on linearizability~\cite{herlihy1990linearizability} when installing checkpoints and roll-back instructions. 
We define the easy integration property more rigorously in Definition~\ref{definition-easy-integration} below. This definition excludes reclamation schemes that alter the plain implementation layout and disallows rolling back into wisely chosen code locations. Consequently, it classifies AOA, FA, NBR and VBR as reclamation schemes that cannot be easily integrated. We state the definition and follow up with more explanations.

\begin{definition} \label{definition-easy-integration}
A reclamation scheme is considered an easily integrated scheme if the following conditions hold:
\begin{enumerate}
    \item \label{guideline-uniform-api} The reclamation scheme is provided as an object. 
    \item \label{guideline-insert-api} The reclamation scheme's API operations may only be inserted in the following code locations: (1) upon the invocation or before the termination of any operation of the plain implementation, (2) as a replacement to \emph{alloc()} and \emph{retire()} calls, or (3) as a replacement to primitive memory access operations.
    \item \label{guideline-same-specification} An API operation that replaces a primitive memory access operation, should be a linearizable implementation of that primitive. 
    \item \label{guideline-well-formed} The integrated implementation should be well-formed.
    \item \label{guideline-node-layout} The reclamation scheme may add new fields to the node's layout and it may access these fields, but it cannot access any other node fields.
\end{enumerate}
\end{definition}

According to Condition~\ref{guideline-uniform-api}, the reclamation scheme should be provided as an object that can be used with all implementations. I.e., as defined in Section~\ref{sec-background}, it should provide a uniform set of API operations, which are the only way to use its mechanism (for more details, see Section~\ref{sec-background}). This Condition also ensures that a reclamation scheme is not adjusted in order to fit specific plain implementations.
Condition~\ref{guideline-insert-api} ensures that the integration procedure is indeed relatively easy (as in EBR and IBR).
Condition~\ref{guideline-same-specification} treats shared memory addresses as objects. I.e., each memory address is an object that provides a set of operations (e.g., read, write, read-modify-write), usually implemented via atomic primitives. According to Condition~\ref{guideline-same-specification}, if a reclamation scheme operation replaces such a primitive, then it should implement it in a linearizable manner. By the locality property of linearizability~\cite{herlihy1990linearizability}, this maintains some level of equivalency between the plain implementation and the integrated one.
Condition~\ref{guideline-well-formed} builds on Condition~\ref{guideline-uniform-api}, and treats the plain implementation and the reclamation scheme implementation as implementations of two separate objects. In particular, the meaning of \emph{well-formed} (as in Section~\ref{sec-background}) in this sense is that the integration cannot move control from within a reclamation-related operation to a point in the plain implementation. Namely, rollbacks from a reclamation API code back into code of the plain implementation are not allowed. 
Finally, Condition~\ref{guideline-node-layout} ensures that the reclamation scheme does not assume anything regarding the node's layout, and does not access any of its original fields. It may only access new fields that it adds to the node.

AOA, NBR and VBR do not satisfy Definition~\ref{definition-easy-integration}, as they do not provide a uniform set of API operations. In particular, inserting roll-back instructions foils Condition~\ref{guideline-well-formed}, as it moves control to an external point in the data-structure code before terminating the current reclamation-related operation execution.
Note that the reclamation API should not include an explicit reclamation operation, as reclamation is expected to occur in the scope of the reclamation scheme code.


\subsection{Wide Applicability} \label{sec-properties-applicability}

In this section we define the applicability of a reclamation scheme to a given plain implementation, and define the wide-applicability property accordingly.
Our applicability definition includes a proper safety engagement, and a reference to the plain implementation's correctness and progress guarantees.
We use linearizability~\cite{herlihy1990linearizability} as our correctness condition, but the definition can be easily adapted to fit other correctness conditions.

\begin{definition} \label{definition-applicability}
We say that a reclamation scheme is applicable to a plain implementation if the following hold:
\begin{enumerate}
    \item {\bf Memory safety:} The reclamation scheme is safe with respect to the plain implementation according to Definition~\ref{definition-smr}.
    \item {\bf Correctness:} The integrated implementation is linearizable\footnote{Note that linearizability (as defined in Section~\ref{sec-background}) refers to the object implemented by the plain implementation, regardless of integration. I.e., the integrated implementation is linearizable with respect to this object even when the reclamation scheme cannot be treated as a separate object (see Section~\ref{sec-properties-easy-integration}).}.
    \item {\bf Progress:} The integrated implementation provides the same progress guarantee as the plain implementation.
\end{enumerate}
\end{definition}

The progress guarantee of a scheme is determined according to~\cite{herlihy2011nature}, and set to the weakest guarantee of any of its operations. For example, Herlihy and Shavit's linked-list is lock-free although its \emph{contains()} operation is wait-free~\cite{herlihy2020art}.

Given the definition of applicability of a reclamation scheme to a plain implementation in Definition~\ref{definition-applicability}, we now define strong applicability and wide applicability of a given SMR. A good property of a reclamation scheme is applicability to as many data structures as possible. 

%
%
The seminal EBR scheme~\cite{harris2001pragmatic,fraser2004practical,brown2015reclaiming} is {\em strongly applicable}, as defined in Definition~\ref{definition-strong-applicability} below (the full proof appears in 
Appendix~\ref{sec-ebr}). 
It is the strongest scheme in terms of applicability. The only assumption it makes with respect to the plain implementation is that \emph{retire()} instructions are properly installed (for more details, see Section~\ref{sec-life-cycle}). 

\begin{definition} \label{definition-strong-applicability}
\textbf{(Strong Applicability)} We say that a reclamation scheme is \emph{strongly applicable} if it is applicable to every plain implementation.
\end{definition}

While strong applicability is highly desirable, to the best of our knowledge, EBR is the only scheme that satisfies it. There are still different extents of applicability for the various reclamation schemes in the literature. We are interested in reclamation schemes that, while not applicable to any imaginable data structure, are still widely applicable to many known data structures. 
To accurately define {\em widely applicable} reclamation schemes, we adopt the definition from previous work~\cite{singh2021nbr}, that defines a large class of well-known and widely-used concurrent data-structure implementations (e.g.,~\cite{harris2001pragmatic,natarajan2014fast,heller2005lazy,ellen2010non,howley2012non}), all applicable to the NBR reclamation scheme.
This class is described in~\cite{singh2021nbr} as containing all data-structure implementations that can be divided into separate interleaving read and write phases. 
\ignore{This seems redundant too...
The motivation behind this separation to phases is to enable efficient reads (as opposed to~\cite{michael2004hazard,wen2018interval,ramalhete2017brief}, reads do not require any reservations), and relatively easy integration of reservations into the code (as all reserved addresses have been written into local pointers during the preceding read phase).} 
We provide the formal definition of this class of data-structure implementations in 
Appendix~\ref{sec-equivalent}. 
We denote such implementations as \emph{access-aware data-structure implementations}, and define wide applicability accordingly:

\begin{definition} \label{definition-wide-applicability}
\textbf{(Wide Applicability)} We say that a reclamation scheme is \emph{widely applicable} if it is applicable to all access-aware data-structure implementations.
\end{definition}

Not all SMRs are widely applicable. We show in 
Appendix~\ref{section-incompatibility} 
that HP, IBR and HE are not widely applicable. 

\section{The ERA Theorem} \label{sec-impossibility}

In Section~\ref{sec-properties} we showed that there exists a reclamation scheme which is both widely applicable and easily integrated (EBR). Assuming that the number of hazard pointers~\cite{michael2004hazard} is asymptotically smaller than the data-structure size (see Section~\ref{sec-properties-robustness}), there also exists a reclamation scheme which is both robust and widely applicable (NBR), and a reclamation scheme which is both robust and easily integrated (HP).
In this section we  prove the main theorem of this paper:

\begin{theorem} \label{theorem-three-properties} 
Any memory reclamation scheme can provide at most two of the following three guarantees: robustness, easy integration and wide applicability.
\end{theorem}

In fact, we prove a stronger result, namely that even weak robustness (see Definition~\ref{definition-weak-robustness}) cannot be achieved when easy integration and wide applicability are provided. This stronger result immediately implies Theorem~\ref{theorem-three-properties}.

In 
Appendix~\ref{section-list-widely-applicable} 
we show that Harris' linked-list is access-aware, and therefore, a widely applicable reclamation scheme must be also  applicable to Harris's linked-list implementation~\cite{harris2001pragmatic}.
For completeness, Harris's plain implementation, including \emph{retire()} calls, is presented in the supplementary material.
The main idea in the proof is to assume in a way of contradiction that an SMR does satisfy all three properties. In this case, it must be applicable to Harris' linked-list, and we then build a specific execution that is not safe for this concurrent linked-list implementation. Thus, no SMR can satisfy all three properties.  

As described in Section~\ref{sec-background}, the list API provides the \emph{insert()}, \emph{delete()} and \emph{contains()} operations, and the nodes comprise of two fields -- an immutable key and a \emph{next} pointer to the node's successor in the list.
The list maintains two sentinel nodes, \emph{head} and \emph{tail}, with the respective $-\infty$ and $\infty$ keys, that are never removed from the list.
Nodes are logically inserted into the list by physically linking them into the list and making them reachable, and are logically deleted from the list by marking their \emph{next} pointer (for more details, see~\cite{harris2001pragmatic,herlihy2020art}). Note that after a node is marked for deletion, it is not necessarily unlinked by the thread that had previously marked it, as it might be unlinked during a concurrent operation. However, the marked node is guaranteed to be unlinked and retired before the \emph{delete()} operation returns.

All three API operations use the \emph{search()} auxiliary method, which is in charge of (1) locating a given key in the list, and (2) unlinking logically deleted (i.e., marked) nodes from the list. 
This method traverses the list (by following \emph{next} pointers) until it finds the first unmarked node with a key greater than or equal to the searched key. After locating such a node, the method might try to physically unlink a sequence of marked nodes from the list. The crucial point here is that marked nodes are not unlinked during the traversal. As opposed to Michael's implementation~\cite{michael2002high} (that was originally designated to fit HP~\cite{michael2004hazard}), when the \emph{search()} method encounters a marked node, it just continues its traversal.

We prove Theorem~\ref{theorem-three-properties} by constructing a specific execution. Let $N\geq 2$ be some fixed constant, and assume $N$ threads are executing Harris's linked list plain implementation, integrated with a widely applicable memory reclamation scheme. 
By Definition~\ref{definition-wide-applicability}, the given reclamation scheme is applicable to this plain implementation. In particular, by Definition~\ref{definition-applicability}, the integrated implementation must provide the same progress guarantee as Harris's algorithm, namely lock-freedom.
Now, assume by contradiction that the scheme is both weakly robust and easily integrated.

\begin{figure}[ht] 
  \centering
  \includegraphics[width=\linewidth,trim= 0in 1.2in 0in 0.5in]{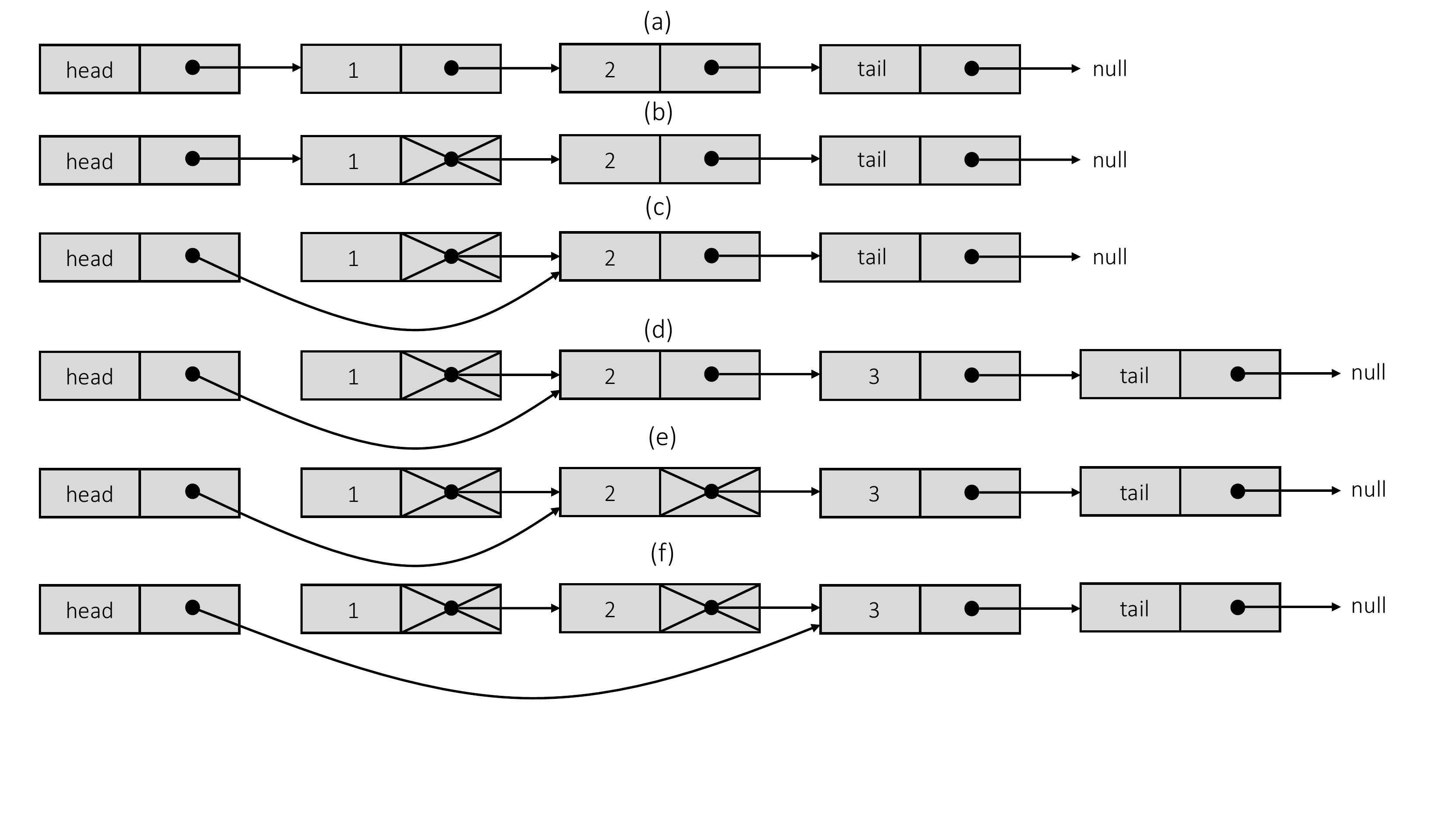}
  \caption{Lower bound illustration ($n=2$).} \label{fig-lb-illustration}
\end{figure}

Initially (stage a in Figure~\ref{fig-lb-illustration}), there are two reachable nodes in the list (besides the $head$ and $tail$ sentinels). Assume that $T_1$ starts executing a \emph{delete(3)} operation. It calls \emph{search(3)}, and starts its traversal by reading \emph{head}'s \emph{next} pointer, which is currently referencing node 1.
At this stage, the scheduler moves control to $T_2$, which executes a \emph{delete(1)} operation. $T_2$ marks node 1 for deletion (stage b) and physically unlinks it from the list (stage c). Next, $T_2$ executes an \emph{insert(3)} operation (stage d) and a \emph{delete(2)} operation (stages e-f). In a similar way, $T_2$ continues calling an alternating sequence of \emph{insert(n+1)} and \emph{delete(n)} (starting from $n=3$). 

For every $n \geq 1$, let $i_n$ be the integer such that $C_{i_n}$ is the configuration after $T_2$ returns from the \emph{delete(n)} execution.
Note that for every $n \geq 2$, after $T_2$ executes \emph{insert(n+1)}, there are four active nodes in the system ($head$, $n$, $n+1$ and $tail$), as the rest of the nodes are already retired before their respective \emph{delete()} operations return. Finally, after $T_2$ executes \emph{delete(n)}, there are three active nodes in the system ($head$, $n+1$ and $tail$), and the nodes $1,\ldots,n$ are already retired. 
Therefore, for every $n \geq 1$, $max\_active_E(i_n) = 4$. 
As the integrated reclamation scheme is weakly robust, there exists a function $f_E$, which is polynomial in $max\_active_E$, such that the number of retired nodes in $C_{i_n}$ is bounded by $f_E(i_n) \cdot N$. 

Let $n > f_E(i_n) \cdot N$ ($n$ must exist as the number of threads $N$ and $max\_active_E$ are constants). 
In $C_{i_n}$, for every $1 \leq i \leq n$, node $i$ has already been retired, and as  $n > f_E(i_n) \cdot N$, by Definition~\ref{definition-weak-robustness}, at least one of the nodes $1,2,\ldots,n$ must already be reclaimed.
In $C_{i_n}$, if a node $i$, $1 \leq i \leq n$, is already retired but not yet reclaimed, then it must be marked and pointing to node $i+1$ via its $next$ pointer. 
To see that the latter is true, recall that, as the scheme is easily integrated, according to Condition~\ref{guideline-node-layout} from Definition~\ref{definition-easy-integration}, the reclamation scheme does not update $next$ pointers. 

Starting from $C_{i_n}$, let the scheduler apply a solo-run by $T_1$. I.e., $T_1$ is the only effective thread in $C_{i_n}$ and on (for more details, see Section~\ref{sec-background}). 
%
$T_1$ continues its traversal from node 1. As all nodes along its path are marked, the traversal's stopping condition does not hold for all nodes along its path (which have keys smaller than $n+1$, and that have not been reclaimed yet), and $T_1$ should continue its traversal. 
By Condition~\ref{guideline-well-formed} from Definition~\ref{definition-easy-integration} (forcing well-formedness), $T_1$ must return from this read operation (as implemented by the reclamation scheme) before it continues its execution. 
As $T_1$ is the only effective thread, and as lock-freedom is guaranteed (see Section~\ref{sec-background}), every such read operation by $T_1$ indeed terminates. 
In addition, according to Condition~\ref{guideline-same-specification} from Definition~\ref{definition-easy-integration}, as long as a node $x$ (for any $1 \leq x \leq n$) is not reclaimed, a read of its \emph{next} pointer by $T_1$ must return a marked reference to the memory (either previously or currently) occupied by node $x+1$.
Therefore, as long as $T_1$ does not encounter a reclaimed node, every read of a $next$ pointer must terminate, returning a (marked) reference to the next node.

Recall that there exists an already reclaimed node $m$ (for some $1 \leq m \leq n$), such that for every $i <m-1$, node $i$'s next pointer is marked, and is pointing to node $i+1$. 
Eventually, $T_1$ must dereference a pointer stored in the memory formerly occupied by $m$. It must assign the content of an invalid $next$ pointer (as $m$ has already been reclaimed) to a local pointer variable, $p$, performing an unsafe memory access by Definition~\ref{definition-safe-access}.
By Definition~\ref{definition-applicability}, the given reclamation scheme is safe with respect to this plain implementation, and therefore, by Definition~\ref{definition-smr}, $T_1$'s local pointer $p$ must be overridden before $T_1$ dereferences it. 
By Condition~\ref{guideline-insert-api} and~\ref{guideline-same-specification} from Definition~\ref{definition-easy-integration}, $T_1$ does not perform any updates before dereferencing $p$, as it contains a marked reference -- a contradiction to the scheme's applicability to the given plain implementation. 
Therefore, a memory reclamation scheme cannot provide robustness, easy integration and wide applicability. \hfill $\blacksquare$


\paragraph{Discussion.}
We proved this result using a specific data structure (Harris's linked-list). This is enough to prove the impossibility in Theorem~\ref{theorem-three-properties}, but it actually provides a stronger result. Even if one tries to achieve robustness and ease of integration only for Harris's linked-list, then this attempt must fail. Therefore, other weaker interpretations of applicability that only require applicability to smaller sets of implementations must still respect the impossibility of Theorem~\ref{theorem-three-properties}, as long as the notion of applicability for an SMR requires it to be applicable to Harris's list (among other implementations). 

An interesting open question is to characterize implementations that are similar to Harris' linked-list. Namely, that memory reclamation with (weak) robustness and ease of integration cannot be obtained for them. 
It is interesting to understand which data structures require special care. 

Finally, we stress the practical importance of this theorem. In order to apply HP to Harris's linked-list, Maged~\cite{michael2002high} modified Harris's implementation to disallow the simultaneous removal of multiple consecutive nodes from the list. While this allowed applying HP to the list, Michael's implementation was slower than the original implementation, see for example the evaluation in~\cite{cohen2015automatic}. Thus, avoiding implementations that are non-trivial for memory reclamation may reduce performance of the concurrent data structures.

\section{Conclusion} \label{sec-conclusion}

In this paper we proposed some theoretical foundation for safe memory reclamation for concurrent data structures. We provided definitions for safe memory reclamation and for three fundamental desirable properties: robustness, easy integration and wide applicability. 
%
We then proved that no reclamation scheme can provide all three desirable properties. I.e., robust reclamation schemes (with limited space overhead) should either be designed for specific implementations, or come  with a relatively complicated manual for proper integration.
Open questions include the formalization of additional interesting properties of SMRs and formal safety proofs for existing SMR schemes.

\ignore{
\begin{table}
  \caption{Frequency of Special Characters}
  \label{tab:freq}
  \begin{tabular}{ccl}
    \toprule
    Non-English or Math&Frequency&Comments\\
    \midrule
    \O & 1 in 1,000& For Swedish names\\
    $\pi$ & 1 in 5& Common in math\\
    \$ & 4 in 5 & Used in business\\
    $\Psi^2_1$ & 1 in 40,000& Unexplained usage\\
  \bottomrule
\end{tabular}
\end{table}

\begin{table*}
  \caption{Some Typical Commands}
  \label{tab:commands}
  \begin{tabular}{ccl}
    \toprule
    Command &A Number & Comments\\
    \midrule
    \texttt{{\char'134}author} & 100& Author \\
    \texttt{{\char'134}table}& 300 & For tables\\
    \texttt{{\char'134}table*}& 400& For wider tables\\
    \bottomrule
  \end{tabular}
\end{table*}
}

\bibliographystyle{ACM-Reference-Format}
\bibliography{refs}

\appendix
\appendix

\section{EBR is Strongly Applicable} \label{sec-ebr}

In this section we prove that the seminal EBR reclamation scheme~\cite{fraser2004practical,harris2001pragmatic,brown2015reclaiming} is strongly applicable. 
According to Definition~\ref{definition-applicability} and~\ref{definition-strong-applicability}, this means that, given any plain implementation: (1) EBR is a safe memory reclamation with respect to that implementation, (2) integrating EBR maintains linearizability, and (3) integrating EBR maintains the same progress guarantee, provided by the plain implementation.

In EBR, the execution is divided into
epochs. The executing threads maintain a shared epoch counter and a shared announcements array.
Upon each invocation of a data-structure operation, the \emph{begin\_op()} operation is called. During the \emph{begin\_op()} operation, the executing thread reads the current global epoch and announces it in the shared announcements array. Then, it checks whether all other threads have announced the current epoch (or a quiescent state). If they have, it increments the shared epoch counter.
Before a data-structure operation returns, the \emph{end\_op()} operation is called, for announcing a quiescent state in the global announcements array.

In addition to the global epoch counter and announcements array, each thread maintains three local retire lists, containing the nodes retired during the last three epochs, respectively (EBR's \emph{retire()} operation only inserts the retired node into the current epoch's retire list).
Once the global epoch counter shows epoch $e$, the retire list for epoch $e-2$ can be reclaimed.

Now, assume a plain data-structure implementation.
We are going to show that EBR is applicable to the given implementation. According to Definition~\ref{definition-applicability}, we need to show that (1) EBR is safe with respect to the given implementation, (2) the integrated implementation is linearizable, and (3) the integrated implementation provide the plain implementation's progress guarantee.

First, we are going to prove that every memory access, during any integrated execution, is safe by Definition~\ref{definition-safe-access}.
Recall that according to Section~\ref{sec-backgound-smr}, the plain implementation already contains \emph{retire()} calls. It is guaranteed that each node is retired at most once, and that nodes are no longer reachable after being retired.
Let $E=C_0,s_1,\ldots$ be an integrated execution, and assume by contradiction that $s_m$ is an unsafe memory access by a thread $T$. W.l.o.g., assume that $s_m$ is the first unsafe memory access during $E$.
By Definition~\ref{definition-safe-access}, $T$ dereferences an invalid pointer, $p$, at $s_m$. Let $s_i$ be the last assignment into $p$ before $s_m$, and let $n$ be the node referenced by $p$ at $C_i$. As $p$ is invalid at $s_m$, and by the choice of $m$, $n$ was reclaimed at some point between $s_i$ and $s_m$.

Let $e$ be $n$'s retire epoch. The global epoch at $s_m$ must be at least $e+2$ (as $n$ is already reclaimed). This means that $T$ must have announced an epoch which is at least $e+1$, upon the invocation of the current operation. As $T$ started its operation during an epoch which is at least $e+1$, $n$ has not been reachable (and obviously, not local to $T$) at any point during the operation execution. By our assumptions in Section~\ref{sec-background}, $T$ cannot have access to $n$'s previously occupied memory address. In particular, its local pointer, $p$, could not have referenced this address -- a contradiction.
Therefore, every memory access, during any integrated execution, is safe by Definition~\ref{definition-safe-access}. By Definition~\ref{definition-smr}, EBR is safe with respect to every plain implementation.

Given that EBR is safe with respect to any given plain implementation, the history of every EBR-integrated execution is equivalent to some history of an execution of the plain implementation. By the definition of linearizability, every EBR-integrated implementation maintains the linearizability of the respective plain implementation.
Finally, EBR maintains any progress guarantee, as its three operations all consist of a finite number of steps.
As EBR is applicable to every plain implementation by Definition~\ref{definition-applicability}, it is strongly applicable by Definition~\ref{definition-strong-applicability}.

\begin{algorithm*}[tbh]
\setlength\multicolsep{0pt}
\small
\begin{multicols}{2}
\begin{algorithmic}[1]

\State private Window *\textbf{search}(key) \label{code-search-call}
\Indent
    \State \textbf{retry}: \textbf{while} (true) \textbf{do} \label{code-search-retry}
    \Indent
        \State Node *pred = head \label{code-search-head}
        \State Node *pred\_next = head $\rightarrow$ next
        \State Node *curr = pred\_next \label{code-search-curr-gets-head-next}
        \State Node *curr\_next = curr $\rightarrow$ next
        \State \textbf{while} (\textbf{isMarked}(curr\_next) || \label{code-search-while-marked}
        \Indent
            \Statex $\; \; \; \; \; \; \; \; \; \; \; \; \; \; \; \; \; \; \; \; \; \; \; \; \; \; \; \; \; \;$ key < curr $\rightarrow$ key) \textbf{do}
            \If{(!isMarked(curr\_next))}
                \State pred = curr
                \State pred\_next = curr\_next
            \EndIf
            \State curr = \textbf{getRef}(curr\_next)
            \If{(curr == tail)} break
            \EndIf
            \State curr\_next = curr $\rightarrow$ next \label{code-search-get-curr-next}
        \EndIndent
        \If{(pred\_next == curr)} \label{code-search-adjacent}
            \If{(\textbf{isMarked}(curr $\rightarrow$ next))} 
                \State \textbf{goto retry}
            \Else  \textbf{ return} new Window(curr, pred) \label{code-search-return-no-cas}
            \EndIf
        \EndIf
        \If{(CAS(\&pred $\rightarrow$ next, pred\_next, curr))} \label{code-search-cas}
            \State Node *tmp = pred\_next
            \State Node *tmp\_next = tmp $\rightarrow$ next
\ignore{
            \While{(tmp $\neq$ curr)}
                \State \textbf{retire}(tmp) \label{code-search-retire}
                \State tmp = \textbf{getRef}(tmp\_next)
                \State tmp\_next = tmp $\rightarrow$ next
            \EndWhile
}
            \If{(\textbf{isMarked}(curr $\rightarrow$ next))} \textbf{goto retry} \label{code-search-retry-after-cas}
            \Else \textbf{ return} new Window(curr, pred) \label{code-search-end}
            \EndIf
        \EndIf
    \EndIndent
\EndIndent

\Statex

\State public boolean \textbf{contains}(key) \label{code-contains-call}
\Indent

        \State Window *window = \textbf{search}(key) \label{code-contains-search}
        \State Node *curr = window $\rightarrow$ curr \label{code-contains-read-curr}
        \State \textbf{return} !\textbf{isMarked}(curr $\rightarrow$ next) $\wedge$ curr $\rightarrow$ key == key \label{code-contains-end}
        
\EndIndent


\Statex
\State public boolean \textbf{insert}(key) \label{code-insert-call}
\Indent
    \State Node *new\_node = \textbf{alloc}(key) \label{code-insert-alloc}
    \While{(true)}
        \State Window *window = \textbf{search}(key) \label{code-insert-search}
        \State Node *pred = window $\rightarrow$ pred
        \State Node *curr = window $\rightarrow$ curr        
        \If{(curr $\rightarrow$ key == key)}
            \State \textbf{retire}(new\_node) \label{code-insert-retire}
            \State \textbf{return} false
        \EndIf
        \State new\_node $\rightarrow$ next = curr
        \If{(CAS(\&pred $\rightarrow$ next, curr, new\_node))} \label{code-insert-cas}
            \State \textbf{return} true \label{code-insert-return-true}
        \EndIf
    \EndWhile
\EndIndent

\Statex
\State public boolean \textbf{delete}(key) \label{code-delete-call}
\Indent
    \While{(true)}
        \State Window *window = \textbf{search}(key) \label{code-delete-search}
        \State Node *pred = window $\rightarrow$ pred
        \State Node *curr = window $\rightarrow$ curr        
        \If{(curr $\rightarrow$ key $\neq$ key)}
            \State \textbf{return} false \label{code-delete-return-false}
        \EndIf
        \State Node *succ = \textbf{getRef}(curr $\rightarrow$ next)
        \State Node *marked = \textbf{getMarked}(succ)
        \If{(!CAS(\&curr $\rightarrow$ next, succ, marked))} \label{code-delete-mark}
            \State \textbf{continue}
        \EndIf
        \If{(!CAS(\&pred $\rightarrow$ next, curr, succ))} \label{code-delete-cas}
            \State \textbf{ search}(key) \label{code-delete-find-physical-deletion}
        \EndIf
        \State \textbf{retire}(curr) \label{code-delete-retire}
        \State \textbf{return} true \label{code-delete-return-true}

    \EndWhile
\EndIndent

\end{algorithmic}
\end{multicols}
\caption{Based on Harris's Non-Blocking Linked-List Implementation}
\label{pseudo-list}
\end{algorithm*}

\section{Harris's Linked-List} \label{section-list}

Harris's plain implementation, including \emph{retire()} calls (see lines~\ref{code-insert-retire} and ~\ref{code-delete-retire}), is presented in Algorithm~\ref{pseudo-list}.
As described in Section~\ref{sec-background}, the list API provides the \emph{contains} (lines~\ref{code-contains-call}-\ref{code-contains-end}), \emph{insert} (lines~\ref{code-insert-call}-\ref{code-insert-return-true}) and \emph{delete} (lines~\ref{code-delete-call}-\ref{code-delete-return-true}) operations, and the nodes comprise of two fields -- an immutable key and a \emph{next} pointer to the node's successor in the list.

The list maintains two sentinel nodes, \emph{head} and \emph{tail}, with the respective $-\infty$ and $\infty$ keys, that are never removed from the list.
Nodes are logically inserted into the list by physically linking them into the list and making them reachable (see line~\ref{code-insert-cas}). Nodes are logically deleted from the list by marking their \emph{next} pointer (for more details, see~\cite{harris2001pragmatic,herlihy2020art}). Note that after a node is marked for deletion in line~\ref{code-delete-mark}, it is not necessarily unlinked by the thread that had previously marked it, as it might be unlinked during a concurrent operation. However, the marked node is guaranteed to be unlinked and retired (in line~\ref{code-delete-retire}) before the operation returns in line~\ref{code-delete-return-true}.

Besides the provided API operations, the implementation uses the \emph{search()} auxiliary method (lines~\ref{code-search-call}-\ref{code-search-end}), which is in charge of (1) locating a given key in the list (the respective node and its predecessor are returned in line~\ref{code-search-return-no-cas} or~\ref{code-search-end}), and (2) unlinking logically deleted nodes from the list (in line~\ref{code-search-cas}).
Marked nodes are not necessarily unlinked during the traversal. The \emph{search()} method tries to unlink a series of marked nodes (line~\ref{code-search-cas}) only if at least one of the output candidates is marked. In such case, it is guaranteed that the condition in line~\ref{code-search-adjacent} does not hold.
\section{Access-Aware Data-Structure Implementations} \label{sec-equivalent}

In this section we formalize the definition of \emph{access-aware data-structure implementations}. This definition relates to the class of implementations, originally defined in~\cite{singh2021nbr}.

As briefly discussed in Section~\ref{sec-properties-applicability}, such implementations can be divided into separate read-only and write phases (the latter type may also include reads). 
Besides these two phase types, each write phase is preceded by a conceptual \emph{reservations} phase. 
The reservation phase should be added during the reclamation scheme integration into the code, and is intended for protecting potential hazardous accesses during the write phase.
Since the reservations phase is not reflected in the given plain implementation (and since some reclamation schemes do not employ protection at all~\cite{DBLP:conf/wdag/SheffiHP21,fraser2004practical,harris2001pragmatic}), we chose to avoid it in our definition. The separation into phases itself is enough for inserting reservation commands before write phases. 

The original definition roughly states two conditions for classifying proper read-only and write phases:
\begin{enumerate}
    \item \label{condition-read} During a read-only phase, shared nodes can be read only if pointers to them were obtained during the current phase.
    \item \label{condition-write} During a write phase, shared nodes can be accessed (either for reads or for writes) only if they were reserved prior to the current phase.
\end{enumerate}

For formalizing the above two conditions, the notion of \emph{obtaining pointers} in condition~\ref{condition-read} should be properly phrased, and an alternative condition for the reservation in condition~\ref{condition-write} should also be provided.

For the latter condition, it suffices to demand that all shared memory accesses during a write phase, do not dereference newly updated pointers.
As long as all dereferenced pointers (either for reads or for writes) are already obtained during the last read-only phase, any integrated reclamation scheme can add the desired reservations phase between the two phases.

The notion of obtaining a pointer can be treated in a similar way to our treatment of safe memory access in Section~\ref{sec:safe-access}.
Given an execution $E=C_0 \cdot s_1 \cdot \ldots$ , a pointer variable $p$, and any configuration $C_m$ in the execution ($m \geq 1$), let $s_i$ be the last update of $p$ in the sub-execution $C_0 \cdot s_1\cdot \ldots C_m$. 
Namely, for $i\leq m$, $p$ is updated in $s_i$, and for every $i<j\leq m$, $s_j$ does not update $p$.
The update of $p$ in $s_i$ may be an allocation of a new node to $p$, a pointer assignment from a global variable (according to Section~\ref{sec-background}, it must be a data-structure entry point), or a dereference of a pointer $q$ (e.g., when reading the \emph{next} pointer of a data-structure node).

Given $j\leq m$, we say that $p$ is $j$-\emph{permitted} at $C_m$, if one of the following holds:
\begin{enumerate}
    \item $s_i$ is an allocation, and at $C_m$, the respective allocated node is still local to the allocating thread.
    \item $s_i$ is a pointer assignment from a global variable, and $j \leq i$.
    \item $s_i$ is a dereference of a pointer $q$, $j < i$, and $q$ is $j$-permitted at $C_{i-1}$.
\end{enumerate}

We are now going to define the class of access-aware data-structure implementations, as follows: a plain implementation is considered as access-aware, if it can be divided into separate alternating read-only and write phases, such that for every derived execution $E=C_0,s_1,\ldots$ and a step $s_m$, it holds that:
\begin{enumerate}
    \item If $s_m$ is a shared-memory read into a local pointer, $p$, during a read-only segment that started in $C_j$ (for some $j\leq m$), then $p$ is $j$-permitted at $C_m$.
    \item If $s_m$ is a shared-memory read into a local pointer, $p$, during a write segment that started in $C_j$ (for some $j\leq m$), then $s_m$ is either an allocation of a new node, a read of a global variable, or a dereference of another pointer variable, $q$, that was $t$-permitted at $C_j$ (given that the last read-only segment began at $C_t$).
    \item If $s_m$ is a shared-memory write, dereferencing a pointer $p$, then the current phase is a write phase. and $p$ was $t$-permitted when the last read-only phase ended (given that the last read-only segment began at $C_t$).
\end{enumerate}

Note that retirements of nodes are not considered as shared memory reads or writes. This definition does not relate to retirement at all.

\section{Harris's Linked-List is an Access-Aware implementation} \label{section-list-widely-applicable}

In this section we are going to prove that Harris's linked-list implementation~\cite{harris2001pragmatic} (as presented in Appendix~\ref{section-list}) is an access-aware data-structure implementation.
This has already been shown in~\cite{singh2021nbr}. However, for the completeness of our main theorem proof (see Section~\ref{sec-impossibility}), we also show it for our definition (from Appendix~\ref{sec-equivalent}). 

We are first going to divide the code into separate read-only and write phases, as follows:

\paragraph{Searches}
The first read-only phase starts in line~\ref{code-search-retry}. 
If the method returns in line~\ref{code-search-return-no-cas}, then a new write phase begins.
If the method does not return in line~\ref{code-search-return-no-cas}, just before line~\ref{code-search-cas} is executed, a new write phase begins, and if the condition in line~\ref{code-search-retry-after-cas} holds, the write phase ends and a new read-only phase begins.
Note that in any case, the \emph{search} method always returns during a write phase. This means that both $pred$ and $curr$ can be assumed to be reserved at this point.

\paragraph{Contains} After returning from the \emph{search} call in line~\ref{code-contains-search}, lines~\ref{code-contains-read-curr}-\ref{code-contains-end} are executed during a write phase.

\paragraph{Inserts} The operation starts accessing shared memory when the \emph{search} auxiliary method is called in line~\ref{code-insert-search}. After it returns, the rest of the loop iteration is executed in a write phase.

\paragraph{Deletes} The operation starts accessing shared memory when the \emph{search} auxiliary method is called in line~\ref{code-delete-search}. After it returns, the rest of the loop iteration is executed in a write phase, unless the \emph{search} method is called again in line~\ref{code-delete-find-physical-deletion}. Then, a new read-only phase is initiated, an so on.

It remains to show that the above separation into phases indeed adheres to the terms defined in Appendix~\ref{sec-equivalent}.
As every read-only phase starts with reading the $head$ global variable in line~\ref{code-search-head} and dereferencing permitted pointers (can be proven by induction), the read-only terms always hold.
In addition, the $pred$ and $curr$ pointers, accessed during the executions of all three operations, are permitted in the last respective read-only-phase.
As the new node, allocated in line~\ref{code-insert-alloc}, is not shared before the execution of line~\ref{code-insert-cas}, accessing it is also permitted. Finally, as mentioned in Appendix~\ref{sec-equivalent}, the retirement in line~\ref{code-delete-retire} is also permitted.
Therefore, Harris's linked-list implementation~\cite{harris2001pragmatic} (as presented in Appendix~\ref{section-list}) is an access-aware data-structure implementation.

\section{Incompatibility to Harris's Linked-List} \label{section-incompatibility}

Many popular reclamation schemes (e.g., HP~\cite{michael2004hazard}, HE~\cite{ramalhete2017brief}, and IBR~\cite{wen2018interval}) are known~\cite{singh2021nbr,kang2020marriage} to not be applicable to the plain implementation, presented in Algorithm~\ref{pseudo-list} (and by the proof from Appendix~\ref{section-list-widely-applicable}, are not widely applicable).
Specifically, HP, IBR, and HE, provide safety via pointers \emph{protection}. I.e., dereferencing is executed in the following manner: the thread first reads the pointer's content, then publishes some data (this data differs between the schemes), and then reads its content again. If the content has not changed between the reads, it is guaranteed that the referenced node is protected, and the thread may safely access it.
Relying on pointers protection is problematic, since a stable pointer value (i.e., the same value is read twice) does not necessarily guarantee safe access to the referenced node.

\begin{figure}[h] 
  \centering
  \includegraphics[width=\linewidth,trim= 0in 2.5in 0in 0in]{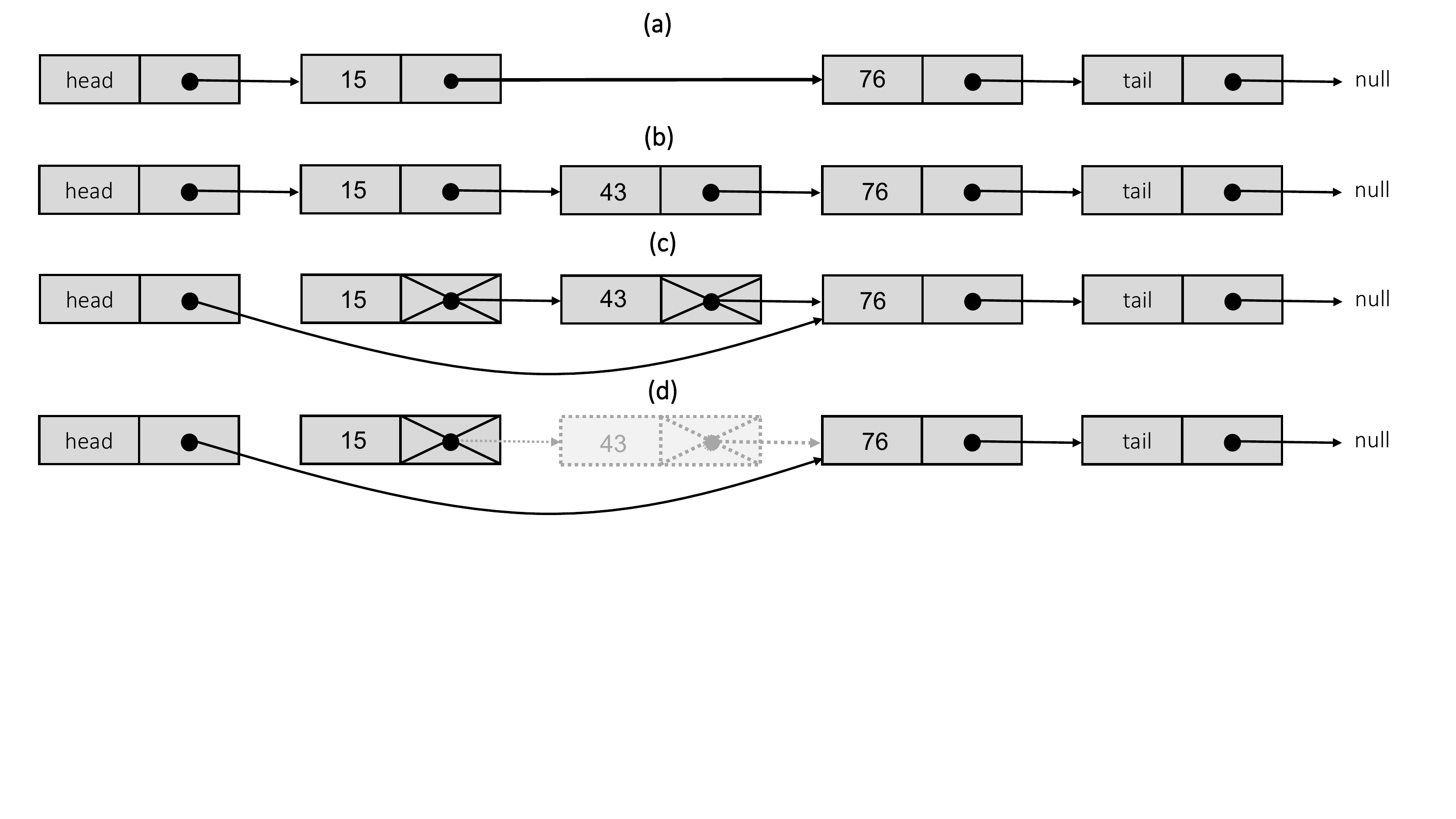}
  \caption{An example for HP's~\cite{michael2004hazard}, HE's~\cite{ramalhete2017brief}, and IBR's~\cite{wen2018interval} limited applicability, using the linked-list implementation from Algorithm~\ref{pseudo-list}.} \label{fig-hp-not-applicable}
\end{figure}

Consider the scenario depicted in Figure~\ref{fig-hp-not-applicable} (the scenario depicted in Figure~\ref{fig-lb-illustration} can also serve as an example). Initially (stage a), the list contains two nodes (15 and 76). At this stage, $T_1$ invokes \emph{insert(58)}, calls the \emph{search()} auxiliary method in line~\ref{code-insert-search}, starts traversing the list, obtains a local pointer to node 15, protects it, and is then halted by the scheduler.
During stage b, some other thread successfully inserts node 43 into the list\footnote{Inserting node 43 after node 15 is protected by $T_1$ is crucial for deriving a contradiction to HE's and IBR's safety. For HP, it can also work if the protection is executed after node 43 is inserted.}.
During stage c, $T_2$ invokes \emph{delete(43)} and $T_3$ invokes \emph{delete(15)}. After both threads find the respective removal window in line~\ref{code-delete-search}, both of them successfully mark their victim node in line~\ref{code-delete-mark}.
At this point, $T_4$ invokes \emph{delete(44)}. During its traversal, it physically unlinks nodes 15 and 43 (by setting $head$'s $next$ reference to point to node 76). Nodes 15 and 43 are eventually retired by $T_3$ and $T_2$ (respectively), before returning from their \emph{delete()} executions (line~\ref{code-delete-retire}). $T_3$ cannot reclaim node 15, as it is protected by $T_1$, but $T_2$ successfully reclaims node 43, as it is not protected by any thread. Eventually, $T_4$ returns in line~\ref{code-delete-return-false}, as there does not exist any node with a 44 key.
After $T_4$ terminates, $T_1$ continues its execution, reads node 15's $next$ pointer, protects node 43's address and re-reads node 15's $next$ pointer. After making sure node 15's $next$ pointer is stable, $T_1$ dereferences its pointer to node 43' address (pre-reclamation), which is already reclaimed. This scenario obviously foils safety, as node 43's memory is either returned to the OS or contains another node at this stage.
Note that even if the memory, previously occupied by node 43, is not returned to the OS, this access foils safety. HP, HE and IBR do not employ a mechanism for dealing with stale values, as depicted in Definition~\ref{definition-smr}.

\end{document}